\documentclass[twocolumn,aps,amsmath,amssymb,pr]{revtex4-1}
\usepackage{bm}
\usepackage{epsfig}
\usepackage{amsmath}
\usepackage{color}
\newcommand{\nix}[1]{}
\usepackage[unicode=true,colorlinks=true,urlcolor=blue,citecolor=blue]{hyperref}
\usepackage{ulem}
\begin{document}

\title{Resonant Excitation of Oscillator with Randomly Shifted Levels} 

\author{A.~P.~Dmitriev}
\email{apd1812@hotmail.com} \affiliation{Ioffe Institute, 194021 St.Petersburg, Russia}

\begin{abstract}
	The problem of resonant excitation of a harmonic oscillator the energy levels of which are slightly shifted under the action of a random potential is solved. It is shown that, in this case, there exists a threshold magnitude of the exciting resonance field, below which the excitation is localized on lower levels, and above which the oscillator is indefinitely excited so that it is necessary to take into account dissipative processes. A method similar to that developed for the oscillator is applied to examine the localization of electrons in a wire with cross-section varying along its length. It is shown, in particular, that there is no localization if this variation is superlinear.
\end{abstract}

\maketitle



    Let us consider a harmonic oscillator with frequency $\omega$, subjected to the action of a random static potential. As an example can serve an electron in a quantum well placed in a quantizing magnetic field, with the role of a random potential played in this case by the potential created by randomly arranged impurities. If the amplitude of this potential is sufficiently small, its influence consists, to a first approximation, in that it randomly shifts the energy levels of the oscillator. The Hamiltonian of an oscillator of this kind can be written as
    \begin{equation}
    H = \hbar \omega \left( \hat{n} +\frac{1}{2} \right) + \varepsilon(\hat{n}),
    \hspace{0.5cm}
    \hat{n} = a^\dagger a,
    \end{equation}
where $\varepsilon(\hat{n})$ is an operator with random eigenvalues  $\varepsilon_n$, with $\langle \varepsilon_n \varepsilon_m \rangle = \varepsilon^2 \delta_{nm}$, $\varepsilon \ll \hbar \omega$. The angular brackets mean the averaging over potential realizations. The eigen functions of the operator, $\hat{n}$, are denoted as $\varphi_n$. 

    Let us assume that an ac field with frequency $\omega$ is applied to the oscillator. Then we have for the wave function of the oscillator the following equation
    \begin{equation}
    \label{eq2}
    i \hbar \frac{d\Psi}{dt} = \hbar \omega \left( \hat{n} +\frac{1}{2} \right) \Psi + \varepsilon(\hat{n}) \Psi + V e^{-i \omega t} a^{\dagger} \Psi + V e^{i \omega t} a \Psi. 
    \end{equation}
Our goal is to find the average probability $W_n$ of oscillator excitation to the  $n$-th level and the average excitation energy $E= \hbar \omega \sum_n n W_n$. For this purpose, we represent the wave function as an expansion in the functions $\varphi_n$:
\begin{equation}
\label{eq3}
\Psi = \sum_n \eta_n(t) e^{-i \omega (n+1/2)t} \varphi_n.
\end{equation}
It is clear that $W_n = \langle |\eta_n|^2 \rangle$. Substituting (\ref{eq3}) in (\ref{eq2}), we obtain the equation for the coefficients $\eta_n$:
\begin{equation}
\label{eq4}
i \hbar \frac{d \eta_n}{dt} = \varepsilon_n \eta_n + V \sqrt{n} \eta_{n-1} 	+ V \sqrt{n+1} \eta_{n+1}. 
\end{equation} 
This equation is a Schr\"odinger equation for a particle moving over sites of a periodic chain with site binding energies $\varepsilon_n$. The difference from the commonly considered case is that the overlapping integral between neighboring sites,  $V \sqrt{n}$, grows with increasing site number as  $\sqrt{n}$, which leads, in particular, to a rise in the "band width" by the same law.

  It is well known that, in the case of the same band width along the chain, a particle with energy not coinciding with the center and band edges is localized in the range with size of the order of mean free path, equal in our designations to $\ell = 4V^2 /\varepsilon^2$, so that $W_n \propto e^{-n/4\ell}$.  It has been stated in a number of reports (see, e.g., \cite{r1} and references therein) that, at the energy corresponding to the band center, there is no localization. It was shown in \cite{r2} that this is not so. A particle is also localized at this energy; however, the exponent contains not the mean free path, but a length $\ell^{\ast}$ that differs from it by a multiplier $\alpha$ on the order of unity, $\ell^{\ast} = \alpha \ell$, i.e., $W_n \propto e^{-n/4\ell^{\ast}}$.

If the mean free path gradually varies along the chain, the exponent contains the integral $ \int_1^n dn'/ \ell(n')$. In our case, $\ell(n) \propto n$ and the band width indefinitely grows with increasing site number. As a consequence, a state with any energy is close to the band center at sufficiently large $n$, and we have
\begin{equation}
\label{eq5}
W_n \propto \exp{\left(- \frac{\varepsilon^2}{16 \alpha V^2} \ln{n}\right)} = n^{-\nu},
\hspace{0.5cm}
\nu = \frac{\varepsilon^2}{16 \alpha V},
\end{equation}
i.e., the excitation localization is of a power-law type, instead of exhibiting an exponential behavior. It follows from (\ref{eq5}) for the average excitation
\begin{equation}
\label{eq6}
E \propto \hbar \omega \sum_n n^{1-\nu} = \hbar \omega \zeta(\nu-1),
\end{equation}
where $\zeta(s)$ is the Riemann zeta function. It can be seen from (\ref{eq6}) that there exists a critical magnitude of the resonance field, $V_c = \varepsilon/4\sqrt{2 \alpha}$, such that at $V>V_c$ the excitation energy indefinitely grows and it is necessary to take into account inelastic processes.

    The above-described phenomenon is, in fact, a particular case of localization in a macroscopically nonuniform one-dimensional system. As one more example we briefly describe the problem of resistance for a wire with cross-section $S(x)$ varying along its length. The resistance of a wire with constant cross-section and conductivity $\sigma$ at zero temperature exponentially grows with its increasing length $L$. This occurs because the wave functions of electrons in the wire are localized along a length 
    $L_c = \hbar \sigma S/e^2$, $\Psi(x) \propto \exp{\left(-x/L_c\right)}$ \cite{r3} (those at length equal to the length of the wire section with resistance of the order $\hbar/e^2$). 
    Suppose now that the wire section increases according to a power law $S = S_0 (1+x/L_0)^q$. 
    Then, instead of $x/L_c$ in the exponent, we will have to write an integral over $x$ and for resistance we get:
\begin{align}
\label{eq7}
&R(L) \propto \exp{\left[\frac{e^2 L_0 \left[(1+L/L_0)^{1-q}-1\right]}{\hbar \sigma S_0 (1-q)}\right]}, \hspace{0.2cm}
q \neq 1;
\notag
\\
&R(L) \propto \left(1+ L/L_0\right)^{e^2 L/\hbar \sigma S_0}, \hspace{0.2cm} q=1.
\end{align}
Consequently, the resistance of a wire grows exponentially with its increasing length at $q < 1$, by a power law at $q = 1$, and tends to a finite limit at $q > 1$, i.e., there is no localization.

\section*{Acknowledgments}

The author is grateful to B.I. Shklovskii for helpful discussions. The work was supported by the Russian Foundation for Basic Research (grant no. 18-02-01016).


\end{document}